\documentclass[%
reprint,
 amsmath,amssymb,
]{revtex4-1}

\usepackage{graphicx}
\usepackage{dcolumn}
\usepackage{bm}

\begin{document}

\preprint{APS/123-QED}

\title{\Large Collisionless periodic orbits in the free-fall three-body problem }

\author{Xiaoming Li$^1$}
\author{Shijun Liao$^{1,2}$}%
\email{sjliao@sjtu.edu.cn}
\affiliation{%
$^1$ Centre of Advanced Computing, School of Naval Architecture, Ocean and Civil Engineering, Shanghai Jiaotong University, China \\
$^2$ Ministry-of-Education Key Laboratory in Scientific and Engineering Computing, Shanghai 200240, China
}%

\date{\today}

\begin{abstract}
Although the free-fall three-body problem have been investigated for more than one century, however, only four collisionless periodic orbits have been found.  In this paper,  we report 234 collisionless periodic orbits of the free-fall three-body system with some mass ratios, including three known collisionless periodic orbits.  Thus,  231 collisionless free-fall periodic orbits among them are entirely new.  In theory,  we can  gain periodic orbits of the free-fall three-body system in arbitrary ratio of mass.  Besides, it is found that, for a given ratio of masses of two bodies,  there exists  a generalized Kepler's third law for the periodic three-body system.  All of these would enrich our knowledge  and deepen our understanding about the famous three-body problem as a whole.
\end{abstract}

\maketitle

\section{Introduction}

The famous three-body problem can be traced back to Newton \cite{Newton1687} in 1680s, and attracted many famous mathematicians and physicists such as  Euler \cite{Euler1767}, Lagrange \cite{Lagrange1772} and so on.  Poincar{\'e} \cite{Poincare1890}  found  that   the  first  integrals for the motion of three-body system  do not exist, and besides
orbits of three-body  system  are  rather  sensitive  to  initial  conditions.  His discovery of the so-called ``sensitivity dependance on initial conditions'' (SDIC)   laid  the foundation of modern chaos theory.  It well explains why in the 300 years only three families of periodic orbits of three-body system were found by Euler \cite{Euler1767} and Lagrange \cite{Lagrange1772}, until 1970s when the Broucke-Hadjidemetriou-H{\'e}non family of periodic orbits were found \citep{Broucke1975,Hadjidemetriou1975a,Hadjidemetriou1975b,Henon1976, Henon1977}.  The famous figure-eight family was numerically discovered by Moore \cite{Moore1993} in 1993 and rediscovered by Chenciner and Montgomery \cite{Chenciner2000} in 2000.   In 2013,  \v{S}uvakov and Dmitra\v{s}inovi\'{c} \cite{Suvakov2013} made a breakthrough to find 13 new distinct periodic orbits by means of numerical methods,  which belong to 11 new families.  In recent years,  periodic orbits of three-body problem have been paid much attention \citep{Iasko2014, Suvakov2014a, Li2017, Li2018}.   In 2017  Li and Liao \cite{Li2017}  found more than six hundred new periodic orbits of three-body system with equal mass, and in 2018  Li et al. \cite{Li2018} further reported more than one thousand new periodic orbits of three-body system with unequal mass, mainly because they can accurately simulate orbits of chaotic three-body systems in a long interval of time  by means of a new numerical strategy, namely the clean numerical simulation (CNS)  \citep{Liao2009, Liao2013-Chaos, Liao2014-3b,Liao2014-SciChina, Liao2015-IJBC, Lin2017}.

The initial conditions of the newly found periodic orbits \citep{Suvakov2013,Iasko2014, Suvakov2014a, Li2017, Li2018} are isosceles collinear configurations.   However, these is another configuration, called the free-fall three-body problem, which have been numerically investigated for more than one century.  The free-fall three-body problem is also called Pythagorean problem,  which was first numerically studied by Burrau \cite{Burrau1913}.    Its first periodic orbit with a binary collision was  found by Szebehely \cite{Szebehely1967}.  Its first collisionless periodic orbit was found by Standish \cite{Standish1970}.  In 1990s Tanikawa et al. \cite{Tanikawa1995} and Tanikawa \cite{Tanikawa2000} reported some  collision orbits of the free-fall three-body problem.
Moeckel et al. \cite{Moeckel2012} proved the existence of periodic brake three-body orbits (i.e., periodic free-fall three-body orbits) with collision in the isosceles configuration.
Tanikawa and Mikkola \cite{Tanikawa2015} located the initial conditions and periods of periodic free-fall three-body orbits with collision.
Yasko and Orlov \cite{Yasko2015} found three collisionless periodic orbits while searching periodic free-fall three-body orbits.
Orlov et al. \cite{Orlov2016} investigated periodic orbits of the free-fall three-body system with equal mass.  In summary, so far, only four collisionless periodic orbits in the free-fall three-body problem have been found.

In this paper, we gain 234 collisionless periodic orbits in the free-fall three-body problem with different mass ratios by means of the the clean numerical simulation (CNS) \citep{Liao2009, Liao2013-Chaos, Liao2014-3b,Liao2014-SciChina, Liao2015-IJBC, Lin2017}, including one periodic orbit found by Standish \cite{Standish1970} and two periodic orbits found by Yasko and Orlov \cite{Yasko2015}. Thus, 231 collisionless free-fall periodic orbits are entirely new.

\section{Numerical approach of searching for periodic orbits}

Let us consider planar three-body system with zero initial velocities and unequal masses.  As shown in Figure~1,  the initial positions of body-1 and body-2 are located at points A $(-0.5,0)$ and  B $(0.5,0)$, respectively, and the initial position of body-3 is at the point C $(x,y)$ in the region $D$ that is surrounded by the $x$ and $y$ axes and a circular segment of unit radius with the point A (-0.5,0) as the centre.  Agekyan and Anosova \cite{Agekyan1968} demonstrated that all possible configurations of the planar free-fall three-body system are included in the region $D$.   For simplicity,  let the Newtonian gravitation constant  be equal to one.

\begin{figure}
\centering
\includegraphics[width=6cm]{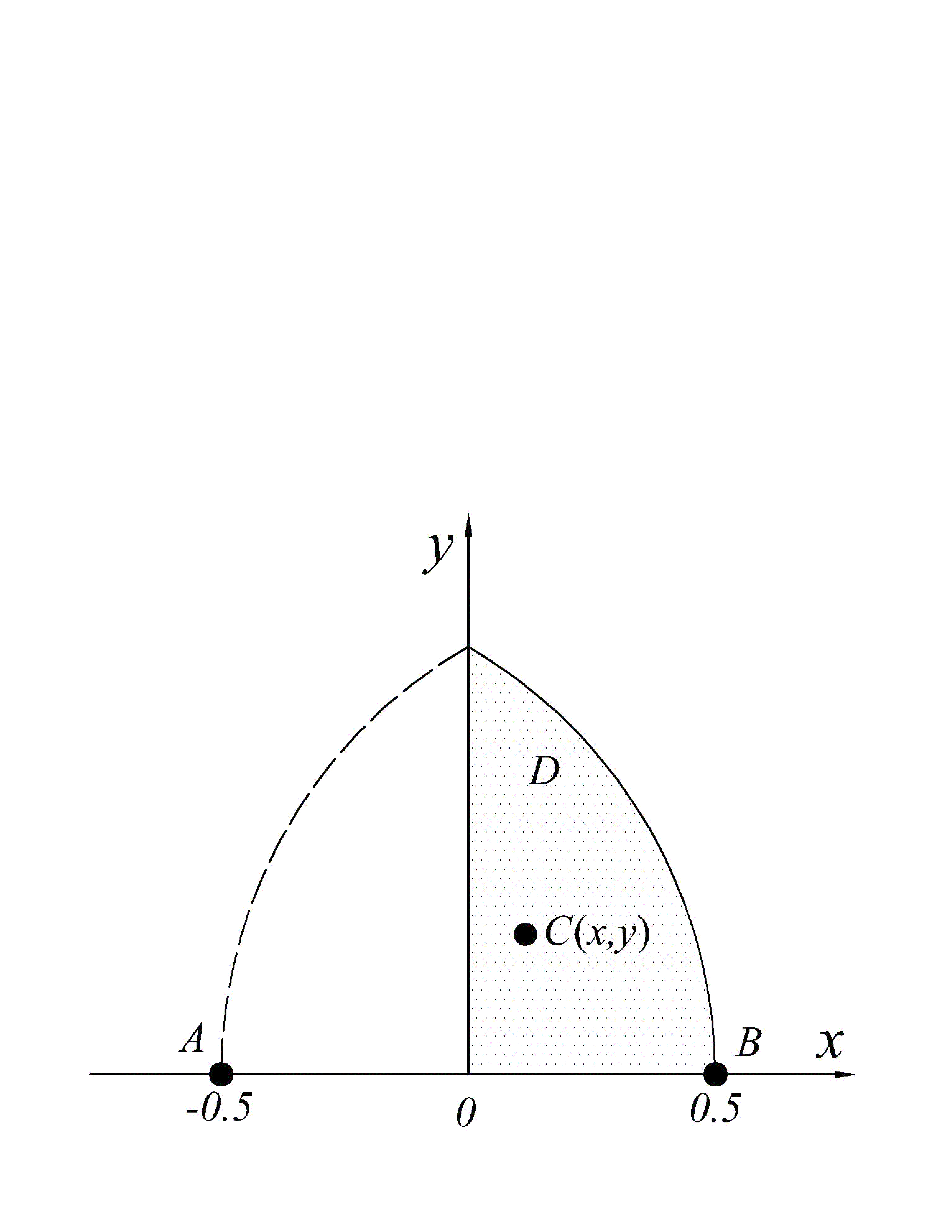}
\caption{\label{fig1}The initial position of three bodies in the configuration space. The initial positions of body-1 and body-2 are located at points A $(-0.5,0)$ and  B $(0.5,0)$, respectively. And the initial position of body-3 is replaced at the point C $(x,y)$ in the region $D$. }
\end{figure}

Because the initial positions of body-1 and body-2 are fixed at point $(-0.5,0)$  and $(0.5,0)$, respectively, the initial position $(x,y)$ of body-3 determines specific solutions of the free-fall three-body system.  Here we denote  $\bm{y}(t)$=($\bm{r}_1(t)$, $\bm{r}_2(t)$, $\bm{r}_3(t)$, $\dot{\bm{r}}_1(t)$, $\dot{\bm{r}}_2(t)$, $\dot{\bm{r}}_3(t)$),  where $\bm{r}_i$ and $\dot{\bm{r}}_i(t)$ are the positions and velocities of the body-$i$ ($i$=1,2,3), respectively.   A periodic solution implies that $\bm{y}(t)$ returns to the initial condition $\bm{y}(t)=\bm{y}(0)$ at time $t=T$, where $T$ is the period.
For specific numerical search of periodic free-fall three-body orbits, the return proximity function is
\begin{equation}
||\bm{y}(t)-\bm{y}(0)|| = \sqrt{\sum_{i=1}^{3}[\bm{r}_i(t)-\bm{r}_i(0)]^2 + \sum_{i=1}^{3}[\dot{\bm{r}}_i(t)-\dot{\bm{r}}_i(0)]^2}
\end{equation}

Here we used the  same strategy as that in our previous discovery of 1bout 2000 new periodic orbits of three-body system \citep{Li2017, Li2018}.   First, we scan the initial positions $(x,y)$ of body-3 in region $D$ with steps of $\Delta x = \Delta y = 0.0001$.  The motion equations of the free-fall triple system are numerically solved by means of the ODE solver dop853 \citep{Hairer1993}.  The approximate initial conditions and periods are gained if the return proximity function is lower than $0.05$. Then,  we correct these approximate initial conditions $(x,y)$ and the periods $T$ by means of the Newton-Raphson method \citep{Farantos1995, Lara2002,  Abad2011}.
 However,  since some periodic three-body orbits might be lost by traditional numerical algorithms in double precision,  we now integrate the motion equations by means of ``clean numerical simulation'' (CNS) \citep{Liao2009, Liao2013-Chaos, Liao2014-3b,Liao2014-SciChina, Liao2015-IJBC, Lin2017}.   The CNS is based on the {\em arbitrary} order of Taylor  series method \citep{Barton1971, Corliss1982,  Barrio2005}  in  {\em arbitrary}  precision \citep{Oyanarte1990, Viswanath2004}, plus a convergence verification using one more simulation with even smaller numerical noises.   A periodic free-fall three-body orbit is obtained when the return proximity function is less than $10^{-6}$.  For more details, please refer to Li and Liao \cite{Li2017} and  Li et al. \cite{Li2018}.

\section{Results}

We focused our numerical search on collisionless periodic orbits in the free-fall three-body problem with period less than 200 and minimum distance between the bodies greater than $10^{-6}$.   Similarly as we did  before \citep{Li2017, Li2018},  we use Montgomery's topological classification method \citep{Montgomery1998,Suvakov2014b} to identify periodic orbits.  The positions of three-body system can be mapped to a point on a unit sphere surface. There are three singular points in the space sphere,  corresponding to the three binary collision in the real space. A collisionless periodic orbit gives a closed cure around three singular points on the shape sphere.  The free group words $a$ and $b$ represent a clockwise turn and counter-clockwise turn around each singular point, respectively. The letters $A$ and $B$ denote the opposite direction of $a$ and $b$, respectively.   The so-called ``free group elements'' of periodic orbits $w$ can be written as $w=w_1w_2w_3\dotsb$, where $w_i$ is any one of $a$, $b$, $A$ and $B$.

We found 30 collisionless periodic orbits in the free-fall three-body system with equal mass ($m_1=m_2= m_3= 1$). The initial conditions and periods of these periodic orbits are listed in Table~\ref{table-A1} in the Appendix A. The free group elements of these periodic orbits are listed in Table~\ref{table-A6} in the Appendix A.
\begin{table}
	\centering
	\caption{The number of periodic orbits with different masses.}
	\label{table1}
	\begin{tabular}{lllc} 
		\hline
$m_1$ \hspace{20pt} & $m_2$ \hspace{20pt} & $m_3$ \hspace{20pt} & number of periodic orbits\\
		\hline
1 & 1 & 1 & 30 \\
1 & 0.8 & 0.8 &29 \\
1 & 0.8 & 0.6 & 69\\
1 & 0.8 & 0.4 & 44\\
1 & 0.8 & 0.2 & 32\\
0.6 & 0.8 & 1 & 30\\
		\hline
	\end{tabular}
\end{table}

\begin{table}
	\centering
	\caption{The initial conditions and periods $T$ of the new collisionless periodic free-fall orbits in the case of  $\bm{r}_1(0)=(-0.5,0)$, $\bm{r}_2(0)=(0.5,0)$, $\bm{r}_3(0)=(x,y)$,  $\dot{\bm{r}}_1(0)=\dot{\bm{r}}_2(0) = \dot{\bm{r}}_3(0)=(0, 0)$ and the Newtonian constant of gravitation $G=1$.}
	\label{table-ini}
	\begin{tabular}{lccr} 
		\hline
$F_{i}(m_1, m_2, m_3)$  & $x$ & $y$  & $T$ \\
		\hline
$F_{10}(1,1,1)$	&	0.3089693008	&	0.4236727692	&	4.8914942162	\\
$F_{1}(1,0.8,0.8)$	&	0.0009114239	&	0.3019805958	&	1.8286248401	\\
$F_{24}(1,0.8,0.8)$	&	0.1527845023	&	0.068494294	&	7.5001089956	\\
$F_{5}(1,0.8,0.6)$	&	0.129088109	&	0.4010761427	&	2.500764871	\\
$F_{27}(1,0.8,0.4)$	&	0.083924021	&	0.3307729197	&	5.3174336486	\\
$F_{6}(1,0.8,0.2)$	&	0.1310631652	&	0.3036588095	&	2.9464698551	\\
		\hline
	\end{tabular}
\end{table}

\begin{figure}
  \centering \includegraphics[scale=0.23]{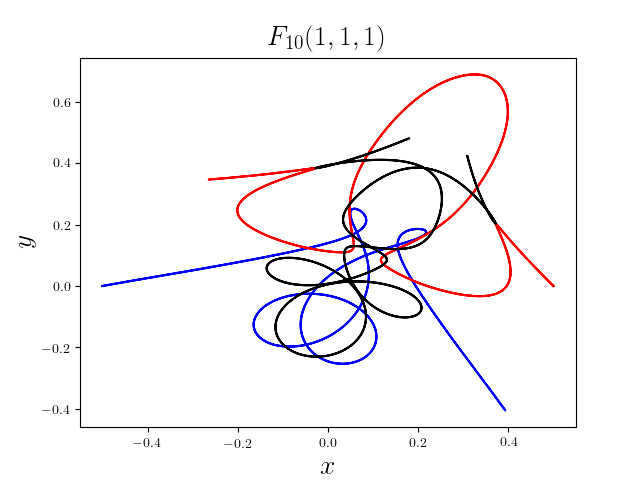}
  \centering \includegraphics[scale=0.23]{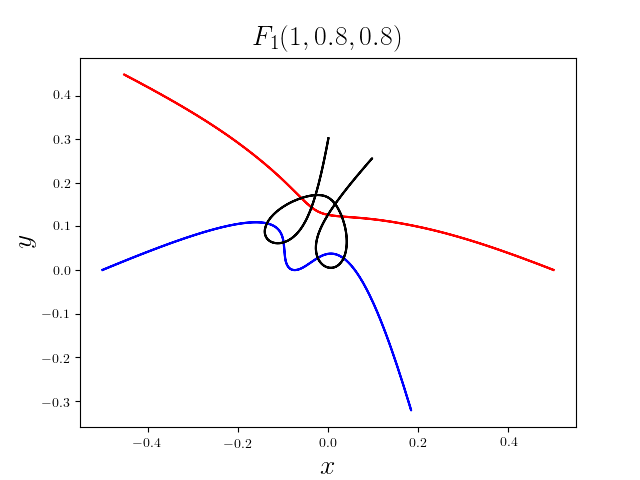}
  \centering \includegraphics[scale=0.23]{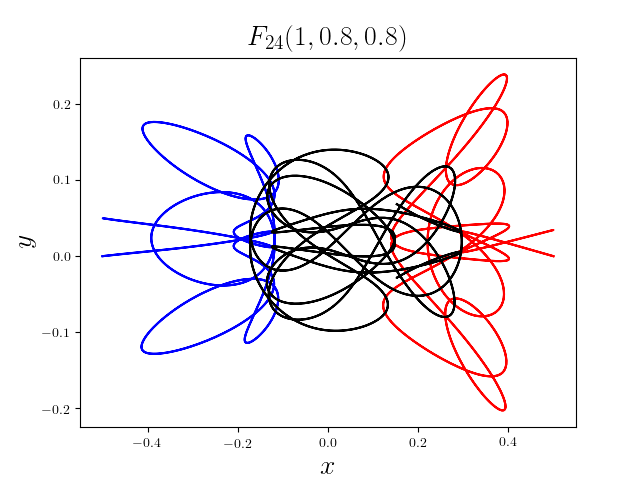}
  \centering \includegraphics[scale=0.23]{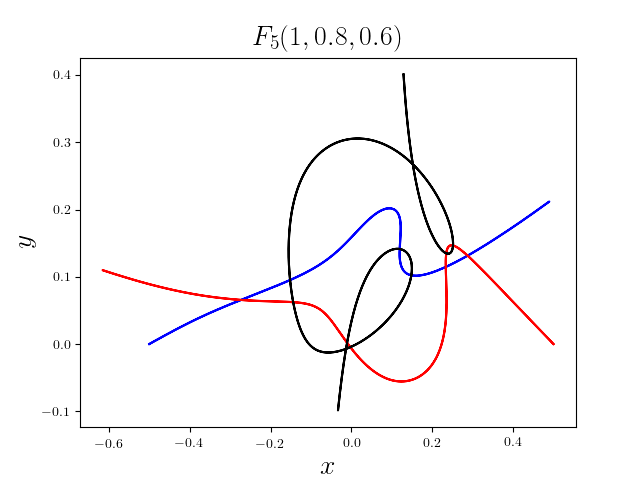}
  \centering \includegraphics[scale=0.23]{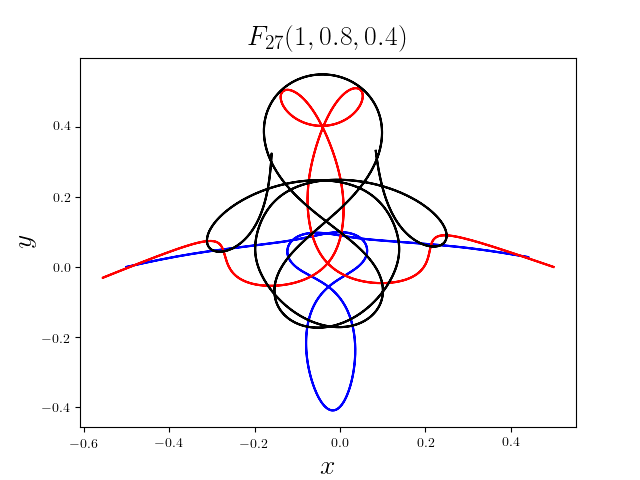}
   \centering \includegraphics[scale=0.23]{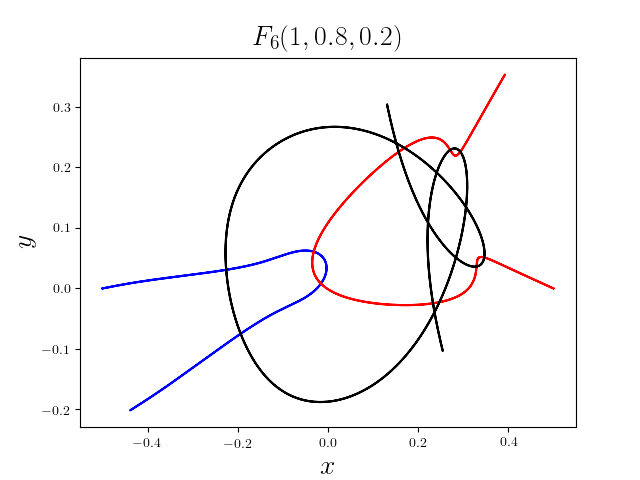}
  \caption{The trajectories of six new collisionless periodic orbits in the free-fall three-body problem. Blue line: body-1,  red line: body-2, black line: body-3.}
  \label{fig2}
\end{figure}

In addition, we also gained collisionless periodic free-fall three-body orbits in some cases of unequal masses, as shown in Table~\ref{table1}.   In summary,  we totally found 234 collisionless periodic orbits in the free-fall three-body problem with different mass ratios. The initial conditions and periods of these periodic orbits are listed in Tables~\ref{table-A1}-\ref{table-A5} in the Appendix A, and their free group elements  are listed in Tables~\ref{table-A6}-\ref{table-A10}. Their movies in the real space and shape sphere are given on the website: \url{http://numericaltank.sjtu.edu.cn/free-fall-3b/free-fall-3b.htm}. The periodic orbit $F_1(0.6,0.8,1)$ was found by Standish \cite{Standish1970}, and the periodic orbits $F_1(1,1,1)$ and $F_2(1,1,1)$ were found by Yasko and Orlov \cite{Yasko2015} (corresponding to their Oribt 16 and 17), respectively.  Except for these three periodic orbits, 231 collisionless periodic orbits are entirely new. Note that one periodic orbit,  named Orbit 19 in Ref. \cite{Yasko2015}, is not found in this paper, because the initial condition of that orbit is not in the region-$D$ considered in this paper.

Although only a few cases of mass ratio listed in Table~\ref{table-ini} are considered here, in theory,  we can similarly gain collisionless  periodic orbits of the free-fall three-body system in {\em arbitrary} ratio of mass.  So, theoretically speaking, there should exist an infinite number of periodic orbits of the free-fall three-body system.

It is found that all bodies of these periodic orbits have zero velocities at time $t=T/2$, where $T$ is the period. After $t=T/2$, the three bodies will go back to the initial positions along the original trajectories. For example, the initial conditions and periods of six newly found collisionless periodic free-fall three-body orbits are listed in Table~\ref{table-ini}, and their trajectories in real space are shown in Fig.~\ref{fig2}.  So, it seems that all periodic trajectories of the free-fall three-body system are {\em not} closed.  This is quite different from the periodic orbits of three-body system with nonzero initial velocities.

\begin{figure}
  \centering \includegraphics[scale=0.23]{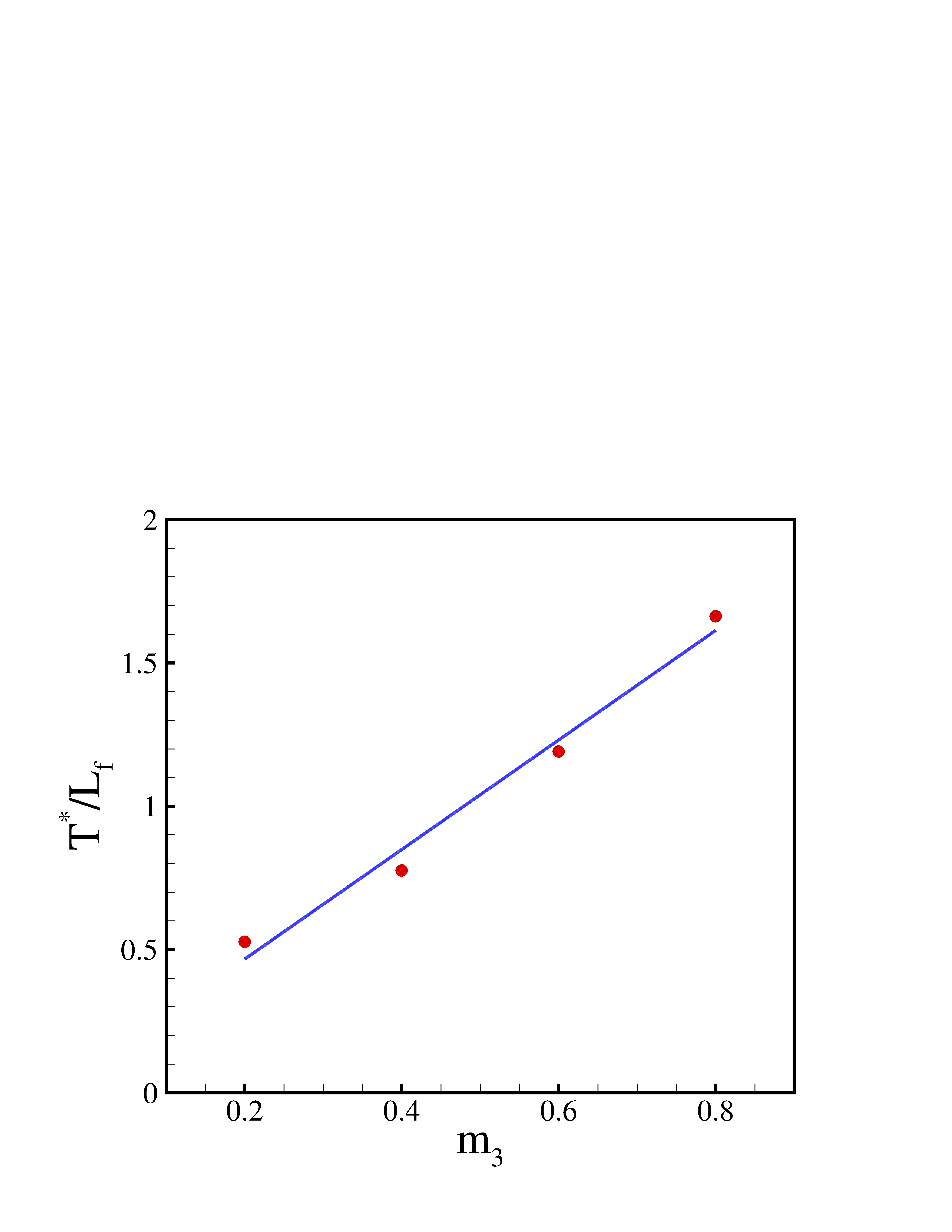}
  \caption{The scale-invariant average period $T^*/L_f=T|E|^{3/2}/L_f$ versus $m_3$ in case of $m_1=1$ and $m_2=0.8$. Symbols: computed results; line: $T^*/L_f= 1.912m_3+0.084$. }
  \label{fig3}
\end{figure}

Li and Liao  \cite{Li2017} found that the periodic orbits with equal mass and initial conditions in case of isosceles collinear configurations have scale-invariant average period $\bar{T}^*=(T/L_f)|E|^{3/2}\approx 2.433$,  where $T$ is the period, $L_f$ is the length of free group element, and $E$ is the total energy of periodic orbits, respectively.  For the collisionless free-fall periodic orbits with equal mass ($m_1=m_2=m_3=1$), we have the scale-invariant average period $\bar{T}^*\approx 2.363$, which is close to $2.433$ with error of 3\%.  It suggests that the scale-invariant average period of the periodic three-body orbits with equal mass ($m_1=m_2=m_3=1$) should be indeed approximately equal to a universal constant for different configurations, say, a generalized Kepler's third law for the periodic three-body system \citep{Dmitrasinovic2015, Li2017, Li2018} should exist in general.
For periodic orbits with unequal mass and initial conditions in case of isosceles collinear configurations, Li et al. \cite{Li2018} found that the scale-invariant average periods increase linearly with $m_3$ in case of $m_1=m_2=1$.  In this paper, for free-fall periodic orbits in case of $m_1=1$ and $m_2=0.8$, the scale-invariant average period is equal to 1.663, 1.191, 0.776 and 0.527 when $m_3$ = 0.8, 0.6, 0.4 and 0.2, respectively.  It is  also  found that the scale-invariant average periods increase linearly with $m_3$, too, as shown in Fig.~\ref{fig3}.   This further suggests that these should indeed exist the so-called generalized Kepler's third law for the periodic three-body system in general.

\section{Conclusions}

Although the free-fall three-body problem have been numerically investigated for more than one century, however, only four collisionless periodic orbits have been found. In this paper,  we report 234 collisionless periodic orbits of the free-fall three-body system with some mass ratios, including one periodic orbit found by Standish \cite{Standish1970} and two periodic orbits found by Yasko and Orlov \cite{Yasko2015}. Thus,  231 periodic orbits among them are entirely new. In theory,  we can similarly gain periodic orbits of the free-fall three-body system in arbitrary ratio of mass.  Besides, it is found that, for a given ratio of masses of two bodies,  there exists a generalized Kepler's third law for the periodic three-body system.    All of these would enrich our knowledge  and deepen our understanding about the famous three-body problem as a whole.

This work was carried out on TH-2 at National Supercomputer Center in Guangzhou, China. It is partly supported by National Natural Science Foundation of China (Approval No. 91752104).

\bibliography{ref}

\setcounter{table}{0}
\renewcommand{\thetable}{A\arabic{table}}
\appendix
\section{The initial conditions, periods and free group elements of periodic orbits}

\begin{table}
	\centering
	\caption{The initial conditions and periods $T$ of the new collisionless periodic free-fall orbits in the case of  $\bm{r}_1(0)=(-0.5,0)$, $\bm{r}_2(0)=(0.5,0)$, $\bm{r}_3(0)=(x,y)$,  $\dot{\bm{r}}_1(0)=\dot{\bm{r}}_2(0) = \dot{\bm{r}}_3(0)=(0, 0)$ and the Newtonian constant of gravitation $G=1$.}
	\label{table-A1}
	\begin{tabular}{lccr} 
		\hline
$F_{i}(m_1, m_2, m_3)$  & $x$ & $y$  & $T$ \\
		\hline
$F_{1}(1,1,1)$	&	0.0207067154	&	0.3133550361	&	2.1740969264	\\
$F_{2}(1,1,1)$	&	0.2053886532	&	0.1952668419	&	1.6896364928	\\
$F_{3}(1,1,1)$	&	0.056266428	&	0.4691503375	&	4.5419125588	\\
$F_{4}(1,1,1)$	&	0.1846729355	&	0.5753740774	&	5.1586391029	\\
$F_{5}(1,1,1)$	&	0.0880412663	&	0.5488924176	&	4.9647695145	\\
$F_{6}(1,1,1)$	&	0.314233405	&	0.5384825297	&	4.8672002993	\\
$F_{7}(1,1,1)$	&	0.0741834378	&	0.5324424488	&	5.4455591108	\\
$F_{8}(1,1,1)$	&	0.2871136862	&	0.5252098584	&	5.1183291764	\\
$F_{9}(1,1,1)$	&	0.0623175679	&	0.4902842102	&	5.1999756815	\\
$F_{10}(1,1,1)$	&	0.3089693008	&	0.4236727692	&	4.8914942162	\\
$F_{11}(1,1,1)$	&	0.4006912162	&	0.3484015305	&	4.2747107264	\\
$F_{12}(1,1,1)$	&	0.3895924236	&	0.3466357955	&	4.2763269534	\\
$F_{13}(1,1,1)$	&	0.1887216677	&	0.5103459122	&	5.9705993269	\\
$F_{14}(1,1,1)$	&	0.1938233757	&	0.6063751102	&	7.6482262218	\\
$F_{15}(1,1,1)$	&	0.2574316821	&	0.363199197	&	6.4040019625	\\
$F_{16}(1,1,1)$	&	0.1472358808	&	0.3114646878	&	6.5387455881	\\
$F_{17}(1,1,1)$	&	0.1403948993	&	0.5079077887	&	8.1068148513	\\
$F_{18}(1,1,1)$	&	0.2104156341	&	0.1080829305	&	5.3952449725	\\
$F_{19}(1,1,1)$	&	0.0410117513	&	0.1882022887	&	8.101047418	\\
$F_{20}(1,1,1)$	&	0.0548189646	&	0.2383162534	&	8.8215019091	\\
$F_{21}(1,1,1)$	&	0.2760396991	&	0.0985020359	&	5.4225040686	\\
$F_{22}(1,1,1)$	&	0.2662035004	&	0.0764101615	&	5.4473716994	\\
$F_{23}(1,1,1)$	&	0.0135939022	&	0.1210361097	&	8.4973950159	\\
$F_{24}(1,1,1)$	&	0.0616197629	&	0.2644961757	&	9.4244325344	\\
$F_{25}(1,1,1)$	&	0.2927953055	&	0.1170062116	&	5.7810404779	\\
$F_{26}(1,1,1)$	&	0.0327160645	&	0.231329184	&	10.0217304818	\\
$F_{27}(1,1,1)$	&	0.1523536847	&	0.1611243738	&	10.3340878156	\\
$F_{28}(1,1,1)$	&	0.0726030609	&	0.3037940331	&	11.8427589787	\\
$F_{29}(1,1,1)$	&	0.3085035769	&	0.1499949186	&	7.293882118	\\
$F_{30}(1,1,1)$	&	0.3401651301	&	0.1168579357	&	6.1839976588	\\
$F_{1}(1,0.8,0.8)$	&	0.0009114239	&	0.3019805958	&	1.8286248401	\\
$F_{2}(1,0.8,0.8)$	&	0.0554096945	&	0.4358787847	&	2.679626506	\\
$F_{3}(1,0.8,0.8)$	&	0.314708305	&	0.3171002382	&	2.5504611351	\\
$F_{4}(1,0.8,0.8)$	&	0.0470065136	&	0.4047349033	&	3.1695717646	\\
$F_{5}(1,0.8,0.8)$	&	0.0428491303	&	0.3894443945	&	3.6573170641	\\
$F_{6}(1,0.8,0.8)$	&	0.1080771428	&	0.5456908479	&	5.5072190289	\\
$F_{7}(1,0.8,0.8)$	&	0.1942960822	&	0.2920607463	&	3.9686010634	\\
$F_{8}(1,0.8,0.8)$	&	0.0403354335	&	0.380070111	&	4.1385691346	\\
$F_{9}(1,0.8,0.8)$	&	0.2016807138	&	0.3445130538	&	4.4274165052	\\
$F_{10}(1,0.8,0.8)$	&	0.1885308478	&	0.3499578547	&	4.9200108461	\\
$F_{11}(1,0.8,0.8)$	&	0.0386388457	&	0.37364164	&	4.6139352019	\\
$F_{12}(1,0.8,0.8)$	&	0.2604939376	&	0.4347691928	&	5.5847931336	\\
$F_{13}(1,0.8,0.8)$	&	0.1947764043	&	0.3822990161	&	5.7130074726	\\
$F_{14}(1,0.8,0.8)$	&	0.1181940386	&	0.5724230518	&	8.1600955849	\\
$F_{15}(1,0.8,0.8)$	&	0.3253474955	&	0.1256079633	&	3.2110917765	\\
$F_{16}(1,0.8,0.8)$	&	0.4292485918	&	0.2660046008	&	4.2565848643	\\
$F_{17}(1,0.8,0.8)$	&	0.0075986006	&	0.1288965886	&	5.1331657257	\\
$F_{18}(1,0.8,0.8)$	&	0.3701389869	&	0.1898186752	&	3.5781033123	\\
$F_{19}(1,0.8,0.8)$	&	0.1691814162	&	0.1508888822	&	5.9395400603	\\
$F_{20}(1,0.8,0.8)$	&	0.2150366646	&	0.4048515918	&	8.5571805321	\\
$F_{21}(1,0.8,0.8)$	&	0.247066563	&	0.0780584462	&	5.1791534733	\\
$F_{22}(1,0.8,0.8)$	&	0.1505569937	&	0.1515667886	&	7.5382540932	\\
$F_{23}(1,0.8,0.8)$	&	0.1674779976	&	0.1427324464	&	7.7585512519	\\
$F_{24}(1,0.8,0.8)$	&	0.1527845023	&	0.068494294	&	7.5001089956	\\
$F_{25}(1,0.8,0.8)$	&	0.1500221997	&	0.0630122343	&	9.2986382763	\\
$F_{26}(1,0.8,0.8)$	&	0.1830123116	&	0.6092070111	&	18.8535978723	\\
$F_{27}(1,0.8,0.8)$	&	0.1464997313	&	0.0473029499	&	11.0720042333	\\
$F_{28}(1,0.8,0.8)$	&	0.1624065612	&	0.138489367	&	13.3138834463	\\
$F_{29}(1,0.8,0.8)$	&	0.1547961371	&	0.1653004219	&	17.5487952899	\\
		\hline
	\end{tabular}
\end{table}

\begin{table}
	\centering
	\caption{The initial conditions and periods $T$ of the new collisionless periodic free-fall orbits in the case of  $\bm{r}_1(0)=(-0.5,0)$, $\bm{r}_2(0)=(0.5,0)$, $\bm{r}_3(0)=(x,y)$,  $\dot{\bm{r}}_1(0)=\dot{\bm{r}}_2(0) = \dot{\bm{r}}_3(0)=(0, 0)$ and the Newtonian constant of gravitation $G=1$.}
	\label{table-A2}
	\begin{tabular}{lccr} 
		\hline
$F_{i}(m_1, m_2, m_3)$  & $x$ & $y$  & $T$ \\
		\hline
$F_{1}(1,0.8,0.6)$	&	0.0445314006	&	0.7402164268	&	2.8030997992	\\
$F_{2}(1,0.8,0.6)$	&	0.0596000878	&	0.7375612563	&	3.4959622385	\\
$F_{3}(1,0.8,0.6)$	&	0.1174556037	&	0.716752919	&	2.7369656923	\\
$F_{4}(1,0.8,0.6)$	&	0.0651532615	&	0.7533349173	&	4.1668179824	\\
$F_{5}(1,0.8,0.6)$	&	0.129088109	&	0.4010761427	&	2.500764871	\\
$F_{6}(1,0.8,0.6)$	&	0.3133248745	&	0.3258575422	&	2.6081696286	\\
$F_{7}(1,0.8,0.6)$	&	0.007896547	&	0.2369793668	&	2.560892572	\\
$F_{8}(1,0.8,0.6)$	&	0.0666851373	&	0.7642377831	&	4.796190851	\\
$F_{9}(1,0.8,0.6)$	&	0.1288237799	&	0.257598905	&	2.563208106	\\
$F_{10}(1,0.8,0.6)$	&	0.1156599683	&	0.4136953992	&	3.1416626327	\\
$F_{11}(1,0.8,0.6)$	&	0.0119593475	&	0.2408362224	&	2.9430485048	\\
$F_{12}(1,0.8,0.6)$	&	0.1023506488	&	0.408866949	&	3.6750954828	\\
$F_{13}(1,0.8,0.6)$	&	0.169463742	&	0.0159900098	&	2.4177885604	\\
$F_{14}(1,0.8,0.6)$	&	0.2707648383	&	0.1015364734	&	2.0709517775	\\
$F_{15}(1,0.8,0.6)$	&	0.361264345	&	0.294301541	&	2.7889383936	\\
$F_{16}(1,0.8,0.6)$	&	0.4108603284	&	0.2242851684	&	2.1055583251	\\
$F_{17}(1,0.8,0.6)$	&	0.0932368548	&	0.4030127578	&	4.1731767189	\\
$F_{18}(1,0.8,0.6)$	&	0.1352566765	&	0.6807175103	&	6.7293787182	\\
$F_{19}(1,0.8,0.6)$	&	0.2725304223	&	0.1732922849	&	2.6175552797	\\
$F_{20}(1,0.8,0.6)$	&	0.3125456686	&	0.086616029	&	1.9950219564	\\
$F_{21}(1,0.8,0.6)$	&	0.1882144839	&	0.3345735824	&	3.5219440458	\\
$F_{22}(1,0.8,0.6)$	&	0.0716145199	&	0.0915187956	&	3.8642156146	\\
$F_{23}(1,0.8,0.6)$	&	0.0869005363	&	0.397993724	&	4.6539481356	\\
$F_{24}(1,0.8,0.6)$	&	0.3563440713	&	0.2981031724	&	3.6559375422	\\
$F_{25}(1,0.8,0.6)$	&	0.2880532139	&	0.4825865222	&	4.6073152572	\\
$F_{26}(1,0.8,0.6)$	&	0.4160772504	&	0.2310953494	&	2.8622355529	\\
$F_{27}(1,0.8,0.6)$	&	0.3987945492	&	0.3025049316	&	3.3736146266	\\
$F_{28}(1,0.8,0.6)$	&	0.077642587	&	0.1746563917	&	4.4628708319	\\
$F_{29}(1,0.8,0.6)$	&	0.0599809897	&	0.0926220002	&	4.3653925159	\\
$F_{30}(1,0.8,0.6)$	&	0.0579903544	&	0.0443882673	&	4.3050365191	\\
$F_{31}(1,0.8,0.6)$	&	0.09346147	&	0.0471665194	&	4.1680025233	\\
$F_{32}(1,0.8,0.6)$	&	0.1325902103	&	0.4123819959	&	5.6679076918	\\
$F_{33}(1,0.8,0.6)$	&	0.104144912	&	0.1482283553	&	4.3424138584	\\
$F_{34}(1,0.8,0.6)$	&	0.1993569698	&	0.0777479367	&	3.8425401785	\\
$F_{35}(1,0.8,0.6)$	&	0.1341720545	&	0.4004224629	&	5.4787823377	\\
$F_{36}(1,0.8,0.6)$	&	0.2974062043	&	0.5279669906	&	6.3977820308	\\
$F_{37}(1,0.8,0.6)$	&	0.2979333584	&	0.4813611322	&	5.0215788607	\\
$F_{38}(1,0.8,0.6)$	&	0.2660209853	&	0.507899546	&	6.132677999	\\
$F_{39}(1,0.8,0.6)$	&	0.352368316	&	0.2982025789	&	4.0622574955	\\
$F_{40}(1,0.8,0.6)$	&	0.4161870623	&	0.232018837	&	3.1969547246	\\
$F_{41}(1,0.8,0.6)$	&	0.4020135877	&	0.3030079422	&	3.6862253065	\\
$F_{42}(1,0.8,0.6)$	&	0.1226012128	&	0.3700294495	&	5.7506652807	\\
$F_{43}(1,0.8,0.6)$	&	0.1005984521	&	0.1891151438	&	4.8651314925	\\
$F_{44}(1,0.8,0.6)$	&	0.2712808805	&	0.1853534625	&	4.0744291974	\\
$F_{45}(1,0.8,0.6)$	&	0.1681172791	&	0.3751181431	&	5.746011321	\\
$F_{46}(1,0.8,0.6)$	&	0.2156072166	&	0.0352712952	&	3.9293028391	\\
$F_{47}(1,0.8,0.6)$	&	0.2256928469	&	0.0765660306	&	3.8667808099	\\
$F_{48}(1,0.8,0.6)$	&	0.1370635146	&	0.4370979031	&	6.4419369834	\\
$F_{49}(1,0.8,0.6)$	&	0.2751838416	&	0.5445094738	&	7.2105813537	\\
$F_{50}(1,0.8,0.6)$	&	0.1171739128	&	0.2424690464	&	5.0831185087	\\
$F_{51}(1,0.8,0.6)$	&	0.1845105186	&	0.4501096494	&	6.2787672531	\\
$F_{52}(1,0.8,0.6)$	&	0.4158959655	&	0.2325668494	&	3.5206991506	\\
$F_{53}(1,0.8,0.6)$	&	0.1631890832	&	0.2103797105	&	5.2589880358	\\
$F_{54}(1,0.8,0.6)$	&	0.1983940001	&	0.4025046079	&	6.3528167889	\\
$F_{55}(1,0.8,0.6)$	&	0.2278308853	&	0.5414697914	&	7.8346766349	\\
$F_{56}(1,0.8,0.6)$	&	0.1400273104	&	0.0081786581	&	4.6831272984	\\
$F_{57}(1,0.8,0.6)$	&	0.2396969281	&	0.6174793436	&	8.4831505812	\\
$F_{58}(1,0.8,0.6)$	&	0.3162910832	&	0.1332526561	&	3.5796429232	\\
$F_{59}(1,0.8,0.6)$	&	0.2092641475	&	0.4289510221	&	6.992742159	\\
		\hline
	\end{tabular}
\end{table}

\begin{table}
	\centering
	\caption{The initial conditions and periods $T$ of the new collisionless periodic free-fall orbits in the case of  $\bm{r}_1(0)=(-0.5,0)$, $\bm{r}_2(0)=(0.5,0)$, $\bm{r}_3(0)=(x,y)$,  $\dot{\bm{r}}_1(0)=\dot{\bm{r}}_2(0) = \dot{\bm{r}}_3(0)=(0, 0)$ and the Newtonian constant of gravitation $G=1$.}
	\label{table-A3}
	\begin{tabular}{lccr} 
		\hline
$F_{i}(m_1, m_2, m_3)$  & $x$ & $y$  & $T$ \\
		\hline
$F_{60}(1,0.8,0.6)$	&	0.2451298945	&	0.5563343875	&	8.6490881257	\\
$F_{61}(1,0.8,0.6)$	&	0.2033682605	&	0.5895584432	&	8.8821315833	\\
$F_{62}(1,0.8,0.6)$	&	0.0792440005	&	0.0978998982	&	6.2066610525	\\
$F_{63}(1,0.8,0.6)$	&	0.1798006367	&	0.5682045433	&	9.3055595206	\\
$F_{64}(1,0.8,0.6)$	&	0.1635781743	&	0.5519695391	&	9.7486824006	\\
$F_{65}(1,0.8,0.6)$	&	0.1496991995	&	0.1853063639	&	7.1021694436	\\
$F_{66}(1,0.8,0.6)$	&	0.1517569566	&	0.5393156268	&	10.2043588748	\\
$F_{67}(1,0.8,0.6)$	&	0.2762432619	&	0.1953715885	&	6.2775182238	\\
$F_{68}(1,0.8,0.6)$	&	0.1707850755	&	0.0856593151	&	6.9891771436	\\
$F_{69}(1,0.8,0.6)$	&	0.1843011931	&	0.2329050114	&	10.3627717531	\\
$F_{1}(1,0.8,0.4)$	&	0.0907814512	&	0.4017862494	&	2.6296980224	\\
$F_{2}(1,0.8,0.4)$	&	0.0731681909	&	0.6154884228	&	3.9909158705	\\
$F_{3}(1,0.8,0.4)$	&	0.1643021746	&	0.3586184694	&	2.8687541956	\\
$F_{4}(1,0.8,0.4)$	&	0.0820569602	&	0.4346030503	&	3.2699876372	\\
$F_{5}(1,0.8,0.4)$	&	0.3039731933	&	0.2570037825	&	2.4050767593	\\
$F_{6}(1,0.8,0.4)$	&	0.003823003	&	0.5336567474	&	3.7553714566	\\
$F_{7}(1,0.8,0.4)$	&	0.2215676344	&	0.3878099512	&	3.3450631915	\\
$F_{8}(1,0.8,0.4)$	&	0.3334492443	&	0.2914625638	&	2.9491668861	\\
$F_{9}(1,0.8,0.4)$	&	0.0734875559	&	0.430759221	&	3.7612586546	\\
$F_{10}(1,0.8,0.4)$	&	0.0041432429	&	0.3565873877	&	3.3430163241	\\
$F_{11}(1,0.8,0.4)$	&	0.364734727	&	0.2936105654	&	2.8392914831	\\
$F_{12}(1,0.8,0.4)$	&	0.3400677332	&	0.3044773669	&	3.4135118241	\\
$F_{13}(1,0.8,0.4)$	&	0.065470114	&	0.3237298022	&	4.9143582571	\\
$F_{14}(1,0.8,0.4)$	&	0.0683822051	&	0.4258256814	&	4.2186879217	\\
$F_{15}(1,0.8,0.4)$	&	0.3051209633	&	0.1717154384	&	2.6362825973	\\
$F_{16}(1,0.8,0.4)$	&	0.381819046	&	0.2165579773	&	2.6549491894	\\
$F_{17}(1,0.8,0.4)$	&	0.3406869343	&	0.2557425196	&	2.9271614571	\\
$F_{18}(1,0.8,0.4)$	&	0.3519249675	&	0.1805278285	&	2.3962764398	\\
$F_{19}(1,0.8,0.4)$	&	0.0067573659	&	0.3551234191	&	3.7006572503	\\
$F_{20}(1,0.8,0.4)$	&	0.3966555518	&	0.2200475447	&	2.5793846625	\\
$F_{21}(1,0.8,0.4)$	&	0.3377184031	&	0.3084466213	&	3.8289301499	\\
$F_{22}(1,0.8,0.4)$	&	0.2559808544	&	0.3985035719	&	4.1088019361	\\
$F_{23}(1,0.8,0.4)$	&	0.3919889126	&	0.2253938155	&	3.0196263432	\\
$F_{24}(1,0.8,0.4)$	&	0.0650272018	&	0.4217643778	&	4.6597662391	\\
$F_{25}(1,0.8,0.4)$	&	0.3507035957	&	0.1514721792	&	2.5093279573	\\
$F_{26}(1,0.8,0.4)$	&	0.0083551748	&	0.3537463108	&	4.0436510812	\\
$F_{27}(1,0.8,0.4)$	&	0.083924021	&	0.3307729197	&	5.3174336486	\\
$F_{28}(1,0.8,0.4)$	&	0.1534813802	&	0.3445489508	&	5.3109746054	\\
$F_{29}(1,0.8,0.4)$	&	0.2639061633	&	0.4003107771	&	4.4615936588	\\
$F_{30}(1,0.8,0.4)$	&	0.3968538929	&	0.2304351014	&	3.36131345	\\
$F_{31}(1,0.8,0.4)$	&	0.3332373486	&	0.3087120049	&	4.2145144195	\\
$F_{32}(1,0.8,0.4)$	&	0.4200988856	&	0.2278457172	&	3.100244422	\\
$F_{33}(1,0.8,0.4)$	&	0.4130847007	&	0.1836206658	&	2.7777999782	\\
$F_{34}(1,0.8,0.4)$	&	0.0626424176	&	0.4185140509	&	5.0903858286	\\
$F_{35}(1,0.8,0.4)$	&	0.2968561297	&	0.1524320976	&	3.0849296562	\\
$F_{36}(1,0.8,0.4)$	&	0.1204686367	&	0.3718569619	&	5.9520762241	\\
$F_{37}(1,0.8,0.4)$	&	0.3388963416	&	0.4809266124	&	6.4469817763	\\
$F_{38}(1,0.8,0.4)$	&	0.3987797585	&	0.2334398833	&	3.6879026872	\\
$F_{39}(1,0.8,0.4)$	&	0.3289050765	&	0.3077086141	&	4.5828783734	\\
$F_{40}(1,0.8,0.4)$	&	0.4211234979	&	0.1897019745	&	3.3348881047	\\
$F_{41}(1,0.8,0.4)$	&	0.1740093467	&	0.4061799018	&	6.945954866	\\
$F_{42}(1,0.8,0.4)$	&	0.4229990097	&	0.1914129869	&	3.6005720033	\\
$F_{43}(1,0.8,0.4)$	&	0.1491465141	&	0.1587305525	&	5.9110461898	\\
$F_{44}(1,0.8,0.4)$	&	0.1071012257	&	0.1862114119	&	7.0200258193	\\
		\hline
	\end{tabular}
\end{table}

\begin{table}
	\centering
	\caption{The initial conditions and periods $T$ of the new collisionless periodic free-fall orbits in the case of  $\bm{r}_1(0)=(-0.5,0)$, $\bm{r}_2(0)=(0.5,0)$, $\bm{r}_3(0)=(x,y)$,  $\dot{\bm{r}}_1(0)=\dot{\bm{r}}_2(0) = \dot{\bm{r}}_3(0)=(0, 0)$ and the Newtonian constant of gravitation $G=1$.}
	\label{table-A4}
	\begin{tabular}{lccr} 
		\hline
$F_{i}(m_1, m_2, m_3)$  & $x$ & $y$  & $T$ \\
		\hline
$F_{1}(1,0.8,0.2)$	&	0.0247367455	&	0.5956711592	&	2.9388503785	\\
$F_{2}(1,0.8,0.2)$	&	0.0788881323	&	0.7268131374	&	3.0188278709	\\
$F_{3}(1,0.8,0.2)$	&	0.0213288594	&	0.6061184155	&	3.434230462	\\
$F_{4}(1,0.8,0.2)$	&	0.0726486463	&	0.3079938956	&	2.7012372544	\\
$F_{5}(1,0.8,0.2)$	&	0.0191076293	&	0.6201117694	&	3.8749764277	\\
$F_{6}(1,0.8,0.2)$	&	0.1310631652	&	0.3036588095	&	2.9464698551	\\
$F_{7}(1,0.8,0.2)$	&	0.0071275432	&	0.4202963121	&	3.229371033	\\
$F_{8}(1,0.8,0.2)$	&	0.0578321783	&	0.4378406225	&	3.3345102032	\\
$F_{9}(1,0.8,0.2)$	&	0.32805122	&	0.4258152565	&	4.5412446489	\\
$F_{10}(1,0.8,0.2)$	&	0.0107850066	&	0.412951734	&	3.515266184	\\
$F_{11}(1,0.8,0.2)$	&	0.0518431819	&	0.4240660566	&	3.6691833705	\\
$F_{12}(1,0.8,0.2)$	&	0.2677953092	&	0.1910152983	&	2.7646135824	\\
$F_{13}(1,0.8,0.2)$	&	0.3185180627	&	0.2611918166	&	3.017887689	\\
$F_{14}(1,0.8,0.2)$	&	0.0161474528	&	0.6441659117	&	4.6837208272	\\
$F_{15}(1,0.8,0.2)$	&	0.0957979866	&	0.7922510102	&	4.8571260165	\\
$F_{16}(1,0.8,0.2)$	&	0.2973457769	&	0.2160446151	&	3.0341335482	\\
$F_{17}(1,0.8,0.2)$	&	0.3192685287	&	0.1435620243	&	2.6213636288	\\
$F_{18}(1,0.8,0.2)$	&	0.0151043114	&	0.6535617498	&	5.0657847438	\\
$F_{19}(1,0.8,0.2)$	&	0.2809348051	&	0.4215890462	&	5.6828932918	\\
$F_{20}(1,0.8,0.2)$	&	0.3505735652	&	0.1838631164	&	2.7916720023	\\
$F_{21}(1,0.8,0.2)$	&	0.0835830692	&	0.3734731952	&	5.6217700487	\\
$F_{22}(1,0.8,0.2)$	&	0.227120654	&	0.3551247364	&	3.8574172516	\\
$F_{23}(1,0.8,0.2)$	&	0.3402615147	&	0.1637450999	&	2.8506619375	\\
$F_{24}(1,0.8,0.2)$	&	0.3793454984	&	0.2079457559	&	3.102833336	\\
$F_{25}(1,0.8,0.2)$	&	0.3188392408	&	0.2307237833	&	3.4919940375	\\
$F_{26}(1,0.8,0.2)$	&	0.2365256725	&	0.3585230897	&	4.1103828594	\\
$F_{27}(1,0.8,0.2)$	&	0.4062681391	&	0.1740808481	&	3.1511604419	\\
$F_{28}(1,0.8,0.2)$	&	0.2121623131	&	0.6190535787	&	9.7560815483	\\
$F_{29}(1,0.8,0.2)$	&	0.4109560735	&	0.1434385455	&	2.9753651036	\\
$F_{30}(1,0.8,0.2)$	&	0.2875973419	&	0.5233120555	&	10.9328732255	\\
$F_{31}(1,0.8,0.2)$	&	0.2145732729	&	0.578482645	&	9.5431888036	\\
$F_{32}(1,0.8,0.2)$	&	0.4217142944	&	0.1509680324	&	3.1849044537	\\
		\hline
	\end{tabular}
\end{table}

\begin{table}
	\centering
	\caption{The initial conditions and periods $T$ of the new collisionless periodic free-fall orbits in the case of  $\bm{r}_1(0)=(-0.5,0)$, $\bm{r}_2(0)=(0.5,0)$, $\bm{r}_3(0)=(x,y)$,  $\dot{\bm{r}}_1(0)=\dot{\bm{r}}_2(0) = \dot{\bm{r}}_3(0)=(0, 0)$ and the Newtonian constant of gravitation $G=1$.}
	\label{table-A5}
	\begin{tabular}{lccr} 
		\hline
$F_{i}(m_1, m_2, m_3)$  & $x$ & $y$  & $T$ \\
		\hline
$F_{1}(0.6,0.8,1)$	&	0.2061730906	&	0.4463433325	&	2.6925803371	\\
$F_{2}(0.6,0.8,1)$	&	0.2632585995	&	0.4021064128	&	2.8891085282	\\
$F_{3}(0.6,0.8,1)$	&	0.3087699287	&	0.281712961	&	2.2526299455	\\
$F_{4}(0.6,0.8,1)$	&	0.3064570616	&	0.2486099306	&	2.3813615017	\\
$F_{5}(0.6,0.8,1)$	&	0.2810833582	&	0.4016924931	&	3.2814736858	\\
$F_{6}(0.6,0.8,1)$	&	0.3226484475	&	0.181377391	&	1.9623770921	\\
$F_{7}(0.6,0.8,1)$	&	0.2687327027	&	0.0903053982	&	1.9434750648	\\
$F_{8}(0.6,0.8,1)$	&	0.323395942	&	0.1697487157	&	2.1207316616	\\
$F_{9}(0.6,0.8,1)$	&	0.3271831808	&	0.2547968571	&	2.62803189	\\
$F_{10}(0.6,0.8,1)$	&	0.289925259	&	0.4030770616	&	3.6787912077	\\
$F_{11}(0.6,0.8,1)$	&	0.342626207	&	0.1830843562	&	2.3182580914	\\
$F_{12}(0.6,0.8,1)$	&	0.3372214876	&	0.260134566	&	2.9032413179	\\
$F_{13}(0.6,0.8,1)$	&	0.333942851	&	0.1166303462	&	1.9513533175	\\
$F_{14}(0.6,0.8,1)$	&	0.3514438378	&	0.1374554522	&	2.133577047	\\
$F_{15}(0.6,0.8,1)$	&	0.3541926333	&	0.1912396242	&	2.5272612515	\\
$F_{16}(0.6,0.8,1)$	&	0.2137677527	&	0.422342585	&	5.8220417242	\\
$F_{17}(0.6,0.8,1)$	&	0.3635347723	&	0.1483383694	&	2.3025570478	\\
$F_{18}(0.6,0.8,1)$	&	0.0697898821	&	0.3190727633	&	6.2973444729	\\
$F_{19}(0.6,0.8,1)$	&	0.3326592654	&	0.2979880796	&	4.4551677666	\\
$F_{20}(0.6,0.8,1)$	&	0.0774214943	&	0.2835443503	&	5.903006977	\\
$F_{21}(0.6,0.8,1)$	&	0.0779210325	&	0.3163212877	&	6.1187532147	\\
$F_{22}(0.6,0.8,1)$	&	0.1428136093	&	0.191027388	&	5.3746485278	\\
$F_{23}(0.6,0.8,1)$	&	0.0894008417	&	0.2215222207	&	6.0165906664	\\
$F_{24}(0.6,0.8,1)$	&	0.041186859	&	0.138693905	&	5.7982241905	\\
$F_{25}(0.6,0.8,1)$	&	0.2645032395	&	0.3738186198	&	6.4584844333	\\
$F_{26}(0.6,0.8,1)$	&	0.0876902685	&	0.1043044639	&	5.9308332463	\\
$F_{27}(0.6,0.8,1)$	&	0.1201933414	&	0.4542911576	&	9.4434059349	\\
$F_{28}(0.6,0.8,1)$	&	0.3461730153	&	0.201836099	&	4.1324912323	\\
$F_{29}(0.6,0.8,1)$	&	0.1715954664	&	0.2463490645	&	8.1331787201	\\
$F_{30}(0.6,0.8,1)$	&	0.1446319096	&	0.4773197126	&	12.5809129586	\\
		\hline
	\end{tabular}
\end{table}

\begin{table}
	\centering
	\caption{The free group elements for the periodic three-body orbits}
	\label{table-A6}
	\begin{tabular}{ll} 
		\hline
$F_{i}(m_1, m_2, m_3)$ & free group element\\
		\hline
$F_{1}(1,1,1)$	&	AbABbaBa	\\
$F_{2}(1,1,1)$	&	ABaBbAba	\\
$F_{3}(1,1,1)$	&	AbAbABaAbaBaBa	\\
$F_{4}(1,1,1)$	&	AbABBaaAAbbaBa	\\
$F_{5}(1,1,1)$	&	AbabABaAbaBABa	\\
$F_{6}(1,1,1)$	&	AbaBAbaABabABa	\\
$F_{7}(1,1,1)$	&	AbaaBAbbBBabAABa	\\
$F_{8}(1,1,1)$	&	AbaBBabAaBAbbABa	\\
$F_{9}(1,1,1)$	&	AbAbABaaAAbaBaBa	\\
$F_{10}(1,1,1)$	&	AbaBBBabAaBAbbbABa	\\
$F_{11}(1,1,1)$	&	AAbABBAbaABabbaBaa	\\
$F_{12}(1,1,1)$	&	AAbABBBaaAAbbbaBaa	\\
$F_{13}(1,1,1)$	&	AbABBBaaBaAbAAbbbaBa	\\
$F_{14}(1,1,1)$	&	AbABAbaBabBAbABabaBa	\\
$F_{15}(1,1,1)$	&	AbaBBAbaBaBAbBabAbABabbABa	\\
$F_{16}(1,1,1)$	&	AbaBBABaaaBAbBabAAAbabbABa	\\
$F_{17}(1,1,1)$	&	AbABBAbaBaBAbBabAbABabbaBa	\\
$F_{18}(1,1,1)$	&	AAbABBBBBBBBAbAAaaBabbbbbbbbaBaa	\\
$F_{19}(1,1,1)$	&	AbABBBBBAbaaaaabAbBaBAAAAABabbbb\\&baBa	\\
$F_{20}(1,1,1)$	&	AbABBBBBAbaaaaaabAbBaBAAAAAABabb\\&bbbaBa	\\
$F_{21}(1,1,1)$	&	AAAbABBBBBBBBBBBAbAAaaBabbbbbbbb\\&bbbaBaaa	\\
$F_{22}(1,1,1)$	&	AAAbABBBBBBBBBBAbAAAaaaBabbbbbbb\\&bbbaBaaa	\\
$F_{23}(1,1,1)$	&	AbABBBBBBBAbaaaaabAbBaBAAAAABabb\\&bbbbbaBa	\\
$F_{24}(1,1,1)$	&	AbABBBBBAbaaaaaaabAbBaBAAAAAAABa\\&bbbbbaBa	\\
$F_{25}(1,1,1)$	&	AAAbABBBBBBBBBBBBAbAAAaaaBabbbbb\\&bbbbbbbaBaaa	\\
$F_{26}(1,1,1)$	&	AbABBBBBBAbaaaaaaabAbbBBaBAAAAAA\\&ABabbbbbbaBa	\\
$F_{27}(1,1,1)$	&	AbaBABAbabABABabaBABAbaBbABababA\\&BAbabaBABababABa	\\
$F_{28}(1,1,1)$	&	AbABBBBBAbaaaaaaaaaaabAbbBBaBAAA\\&AAAAAAAABabbbbbaBa	\\
$F_{29}(1,1,1)$	&	AAAbABBBBBBBBBBBBBBBBBBAbAAAaaaB\\&abbbbbbbbbbbbbbbbbbaBaaa	\\
$F_{30}(1,1,1)$	&	AAAAbABBBBBBBBBBBBBBBBBAbAAAAaaa\\&aBabbbbbbbbbbbbbbbbbaBaaaa	\\
$F_{1}(1,0.8,0.8)$	&	BaBbAb	\\
$F_{2}(1,0.8,0.8)$	&	AbaBbABa	\\
$F_{3}(1,0.8,0.8)$	&	AAbABbaBaa	\\
$F_{4}(1,0.8,0.8)$	&	AbaBBbbABa	\\
$F_{5}(1,0.8,0.8)$	&	AbABBBbbbaBa	\\
$F_{6}(1,0.8,0.8)$	&	AbaBAbaABabABa	\\
$F_{7}(1,0.8,0.8)$	&	AbaBAbbBBabABa	\\
$F_{8}(1,0.8,0.8)$	&	AbABBBBbbbbaBa	\\
$F_{9}(1,0.8,0.8)$	&	AbaBBabAaBAbbABa	\\
$F_{10}(1,0.8,0.8)$	&	AbaBBAbaABabbABa	\\
$F_{11}(1,0.8,0.8)$	&	AbABBBBBbbbbbaBa	\\
$F_{12}(1,0.8,0.8)$	&	AbaBBBabAaBAbbbABa	\\
$F_{13}(1,0.8,0.8)$	&	AbaBBBAbaABabbbABa	\\
$F_{14}(1,0.8,0.8)$	&	AbaBAbaBBaAbbABabABa	\\
$F_{15}(1,0.8,0.8)$	&	AAAbaBBAbaABabbABaaa	\\
$F_{16}(1,0.8,0.8)$	&	AAAbABBAbaABabbaBaaa	\\
$F_{17}(1,0.8,0.8)$	&	BaBAAAABaBBbbAbaaaabAb	\\
$F_{18}(1,0.8,0.8)$	&	AAAbaBBBabAaBAbbbABaaa	\\
$F_{19}(1,0.8,0.8)$	&	AbaBABababABaAbaBABAbabABa	\\
$F_{20}(1,0.8,0.8)$	&	AbaBBBAbbAbABaAbaBaBBabbbABa	\\

		\hline
	\end{tabular}
\end{table}

\begin{table}
	\centering
	\caption{The free group elements for the periodic three-body orbits}
	\label{table-A7}
	\begin{tabular}{ll} 
		\hline
$F_{i}(m_1, m_2, m_3)$ & free group element\\
		\hline
$F_{21}(1,0.8,0.8)$	&	AAbaBBAbabaBAbBabABABabbABaa	\\
$F_{22}(1,0.8,0.8)$	&	AbaBABababABBabAaBAbbaBABAbabABa	\\
$F_{23}(1,0.8,0.8)$	&	AbaBABababABABabAaBAbabaBABAbabA\\&Ba	\\
$F_{24}(1,0.8,0.8)$	&	AbaBABAbabABABabAaBAbabaBABababA\\&Ba	\\
$F_{25}(1,0.8,0.8)$	&	AbaBABAbabABABababABaAbaBABAbaba\\&BABababABa	\\
$F_{26}(1,0.8,0.8)$	&	AbaBAbbAbaBBabAbaBBaBAbaBbABabAb\\&bABaBAbbABBBabABa	\\
$F_{27}(1,0.8,0.8)$	&	AbaBABAbabaBABababABABabAaBAbaba\\&BABAbabABABababABa	\\
$F_{28}(1,0.8,0.8)$	&	AbaBABababABABaBABAbabaBABabAaBA\\&babABABababAbabaBABAbabABa	\\
$F_{29}(1,0.8,0.8)$	&	AbaBABababAbabaBABababABAbaBABAb\\&abABaAbaBABababABabaBABAbabABABa\\&BABAbabABa	\\
$F_{1}(1,0.8,0.6)$	&	AbBa	\\
$F_{2}(1,0.8,0.6)$	&	AbaABa	\\
$F_{3}(1,0.8,0.6)$	&	AbAaBa	\\
$F_{4}(1,0.8,0.6)$	&	AbaaAABa	\\
$F_{5}(1,0.8,0.6)$	&	AbAbBaBa	\\
$F_{6}(1,0.8,0.6)$	&	AAbaABaa	\\
$F_{7}(1,0.8,0.6)$	&	BaBaaAAbAb	\\
$F_{8}(1,0.8,0.6)$	&	AbaaaAAABa	\\
$F_{9}(1,0.8,0.6)$	&	AbAbaABaBa	\\
$F_{10}(1,0.8,0.6)$	&	AbAbbBBaBa	\\
$F_{11}(1,0.8,0.6)$	&	BaBaaaAAAbAb	\\
$F_{12}(1,0.8,0.6)$	&	AbAbbbBBBaBa	\\
$F_{13}(1,0.8,0.6)$	&	ABaBaaAAbAba	\\
$F_{14}(1,0.8,0.6)$	&	AABaBaAbAbaa	\\
$F_{15}(1,0.8,0.6)$	&	AAbAbbBBaBaa	\\
$F_{16}(1,0.8,0.6)$	&	AAAbAbBaBaaaA	\\
$F_{17}(1,0.8,0.6)$	&	AbAbbbbBBBBaBa	\\
$F_{18}(1,0.8,0.6)$	&	AbABAbbBBabaBa	\\
$F_{19}(1,0.8,0.6)$	&	AAbAbaaAABaBaa	\\
$F_{20}(1,0.8,0.6)$	&	AAABaBaAbAbaaa	\\
$F_{21}(1,0.8,0.6)$	&	AbABaaaAAAbaBa	\\
$F_{22}(1,0.8,0.6)$	&	AbAbAbAbBaBaBaBa	\\
$F_{23}(1,0.8,0.6)$	&	AbAbbbbbBBBBBaBa	\\
$F_{24}(1,0.8,0.6)$	&	AAbAbbbbBBBBaBaa	\\
$F_{25}(1,0.8,0.6)$	&	AbAbaaaaAAAABaBa	\\
$F_{26}(1,0.8,0.6)$	&	AAAbAbbbBBBaBaaa	\\
$F_{27}(1,0.8,0.6)$	&	AAbAbaaaAAABaBaaA	\\
$F_{28}(1,0.8,0.6)$	&	AbAbabABaAbaBABaBa	\\
$F_{29}(1,0.8,0.6)$	&	AbAbbaBAbBabABBaBa	\\
$F_{30}(1,0.8,0.6)$	&	AbAbbabAbBaBABBaBa	\\
$F_{31}(1,0.8,0.6)$	&	AbaBABAbAaBababABa	\\
$F_{32}(1,0.8,0.6)$	&	AbAbBaBAbBabAbBaBa	\\
$F_{33}(1,0.8,0.6)$	&	AbaBABabAaBAbabABa	\\
$F_{34}(1,0.8,0.6)$	&	AAbAbAbaBbABaBaBaa	\\
$F_{35}(1,0.8,0.6)$	&	AbAbAbABaAbaBaBaBa	\\
$F_{36}(1,0.8,0.6)$	&	AbABBAbAbBaBabbaBa	\\
$F_{37}(1,0.8,0.6)$	&	AbAbaaaaaAAAAABaBa	\\
$F_{38}(1,0.8,0.6)$	&	AbABABaBaAbAbabaBa	\\
$F_{39}(1,0.8,0.6)$	&	AAbAbbbbbBBBBBaBaa	\\
$F_{40}(1,0.8,0.6)$	&	AAAbAbbbbBBBBaBaaaA	\\
$F_{41}(1,0.8,0.6)$	&	AAbAbaaaaAAAABaBaaA	\\
$F_{42}(1,0.8,0.6)$	&	AbAbabABaaAAbaBABaBa	\\
$F_{43}(1,0.8,0.6)$	&	AbABABaBaaAAbAbabaBa	\\
$F_{44}(1,0.8,0.6)$	&	AAbAbabAbAaBaBABaBaa	\\
$F_{45}(1,0.8,0.6)$	&	AbAbaabAbAaBaBAABaBa	\\
	\hline
	\end{tabular}
\end{table}

\begin{table}
	\centering
	\caption{The free group elements for the periodic three-body orbits}
	\label{table-A8}
	\begin{tabular}{ll} 
		\hline
$F_{i}(m_1, m_2, m_3)$ & free group element\\
		\hline
$F_{46}(1,0.8,0.6)$	&	AAbAbbaBAbBabABBaBaa	\\
$F_{47}(1,0.8,0.6)$	&	AAbAbabABaAbaBABaBaa	\\
$F_{48}(1,0.8,0.6)$	&	AbAbBBaBAbBabAbbBaBa	\\
$F_{49}(1,0.8,0.6)$	&	AbABBBAbAbBaBabbbaBa	\\
$F_{50}(1,0.8,0.6)$	&	AbAbaabABaAbaBAABaBa	\\
$F_{51}(1,0.8,0.6)$	&	AbAbaAbAbaABaBaABaBa	\\
$F_{52}(1,0.8,0.6)$	&	AAAbAbbbbbBBBBBaBaaaA	\\
$F_{53}(1,0.8,0.6)$	&	AbaBabaBabAaBAbABAbABa	\\
$F_{54}(1,0.8,0.6)$	&	AbAbaaabAbAaBaBAAABaBa	\\
$F_{55}(1,0.8,0.6)$	&	AbABBBBAbAbBaBabbbbaBa	\\
$F_{56}(1,0.8,0.6)$	&	AbaBAABaBaaAAbAbaabABa	\\
$F_{57}(1,0.8,0.6)$	&	AbABBBBBAbAaBabbbbbaBa	\\
$F_{58}(1,0.8,0.6)$	&	AAAbAbabAbAaBaBABaBaaa	\\
$F_{59}(1,0.8,0.6)$	&	AbAbaaaabAbAaBaBAAAABaBa	\\
$F_{60}(1,0.8,0.6)$	&	AbABBBBAbAbbBBaBabbbbaBa	\\
$F_{61}(1,0.8,0.6)$	&	AbABBBBBBAbAaBabbbbbbaBa	\\
$F_{62}(1,0.8,0.6)$	&	AbABABaBABAbAaBababAbabaBa	\\
$F_{63}(1,0.8,0.6)$	&	AbABBBBBBBAbAaBabbbbbbbaBa	\\
$F_{64}(1,0.8,0.6)$	&	AbABBBBBBBBAbAaBabbbbbbbbaBa	\\
$F_{65}(1,0.8,0.6)$	&	AbaBAbaBabABabAaBAbaBAbABabABa	\\
$F_{66}(1,0.8,0.6)$	&	AbABBBBBBBBBAbAaBabbbbbbbbbaBa	\\
$F_{67}(1,0.8,0.6)$	&	AAbAbabAbabAbaBbABaBABaBABaBaa	\\
$F_{68}(1,0.8,0.6)$	&	ABaBabaBabbbaBaBbAbABBBAbABAbAba	\\
$F_{69}(1,0.8,0.6)$	&	AbaBabABabABAbABBBAbAaBabbbaBaba\\&BAbaBAbABa	\\
$F_{1}(1,0.8,0.4)$	&	AbAbBaBa	\\
$F_{2}(1,0.8,0.4)$	&	AbABbaBa	\\
$F_{3}(1,0.8,0.4)$	&	AbAbAaBaBa	\\
$F_{4}(1,0.8,0.4)$	&	AbAbbBBaBa	\\
$F_{5}(1,0.8,0.4)$	&	AAbAbBaBaa	\\
$F_{6}(1,0.8,0.4)$	&	BaBBBbbbAb	\\
$F_{7}(1,0.8,0.4)$	&	AbAbAAaaBaBa	\\
$F_{8}(1,0.8,0.4)$	&	AAbAbbBBaBaa	\\
$F_{9}(1,0.8,0.4)$	&	AbAbbbBBBaBa	\\
$F_{10}(1,0.8,0.4)$	&	BaBaaaAAAbAb	\\
$F_{11}(1,0.8,0.4)$	&	AAbAbaABaBaa	\\
$F_{12}(1,0.8,0.4)$	&	AAbAbbbBBBaBaa	\\
$F_{13}(1,0.8,0.4)$	&	AbAbbaBbABBaBa	\\
$F_{14}(1,0.8,0.4)$	&	AbAbbbbBBBBaBa	\\
$F_{15}(1,0.8,0.4)$	&	AABaBaaAAbAbaa	\\
$F_{16}(1,0.8,0.4)$	&	AAAbAbbBBaBaaa	\\
$F_{17}(1,0.8,0.4)$	&	AAbAbAAaaBaBaa	\\
$F_{18}(1,0.8,0.4)$	&	AAAbAbAaBaBaaa	\\
$F_{19}(1,0.8,0.4)$	&	BaBaaaaAAAAbAb	\\
$F_{20}(1,0.8,0.4)$	&	AAAbAbaABaBaaa	\\
$F_{21}(1,0.8,0.4)$	&	AAbAbbbbBBBBaBaa	\\
$F_{22}(1,0.8,0.4)$	&	AbAbAAAAaaaaBaBa	\\
$F_{23}(1,0.8,0.4)$	&	AAAbAbbbBBBaBaaa	\\
$F_{24}(1,0.8,0.4)$	&	AbAbbbbbBBBBBaBa	\\
$F_{25}(1,0.8,0.4)$	&	AAABaBaaAAbAbaaa	\\
$F_{26}(1,0.8,0.4)$	&	BaBaaaaaAAAAAbAb	\\
$F_{27}(1,0.8,0.4)$	&	AbAbabAbAaBaBABaBa	\\
$F_{28}(1,0.8,0.4)$	&	AbAbABAbAaBabaBaBa	\\
$F_{29}(1,0.8,0.4)$	&	AbAbAAAAAaaaaaBaBa	\\
$F_{30}(1,0.8,0.4)$	&	AAAbAbbbbBBBBaBaaa	\\
$F_{31}(1,0.8,0.4)$	&	AAbAbbbbbBBBBBaBaa	\\
$F_{32}(1,0.8,0.4)$	&	AAAbAbaaaAAABaBaaa	\\
$F_{33}(1,0.8,0.4)$	&	AAAAbAbbbBBBaBaaaa	\\
$F_{34}(1,0.8,0.4)$	&	AbAbbbbbbBBBBBBaBa	\\
$F_{35}(1,0.8,0.4)$	&	AABaBaaaaAAAAbAbaa	\\
$F_{36}(1,0.8,0.4)$	&	AbAbaabAbAaBaBAABaBa	\\

		\hline
	\end{tabular}
\end{table}

\begin{table}
	\centering
	\caption{The free group elements for the periodic three-body orbits}
	\label{table-A9}
	\begin{tabular}{ll} 
		\hline
$F_{i}(m_1, m_2, m_3)$ & free group element\\
		\hline
$F_{37}(1,0.8,0.4)$	&	AbABAbAbAAaaBaBabaBa	\\
$F_{38}(1,0.8,0.4)$	&	AAAbAbbbbbBBBBBaBaaa	\\
$F_{39}(1,0.8,0.4)$	&	AAbAbbbbbbBBBBBBaBaa	\\
$F_{40}(1,0.8,0.4)$	&	AAAAbAbbbbbBBBBBaBaaaa	\\
$F_{41}(1,0.8,0.4)$	&	AbAbaaaabAbAaBaBAAAABaBa	\\
$F_{42}(1,0.8,0.4)$	&	AAAAbAbbbbbbBBBBBBaBaaaa	\\
$F_{43}(1,0.8,0.4)$	&	AbaBaBaBaBabAaBAbAbAbAbABa	\\
$F_{44}(1,0.8,0.4)$	&	AbAbAbAbAbAbAbAaBaBaBaBaBaBaBa	\\
$F_{1}(1,0.8,0.2)$	&	BaAb	\\
$F_{2}(1,0.8,0.2)$	&	AbAaBa	\\
$F_{3}(1,0.8,0.2)$	&	BabBAb	\\
$F_{4}(1,0.8,0.2)$	&	AbAbBaBa	\\
$F_{5}(1,0.8,0.2)$	&	BabbBBAb	\\
$F_{6}(1,0.8,0.2)$	&	AbAbAaBaBa	\\
$F_{7}(1,0.8,0.2)$	&	BaBaaAAbAb	\\
$F_{8}(1,0.8,0.2)$	&	AbAbbBBaBa	\\
$F_{9}(1,0.8,0.2)$	&	AbABaAbaBa	\\
$F_{10}(1,0.8,0.2)$	&	BaBaaaAAAbAb	\\
$F_{11}(1,0.8,0.2)$	&	AbAbbbBBBaBa	\\
$F_{12}(1,0.8,0.2)$	&	AAbAbAaBaBaa	\\
$F_{13}(1,0.8,0.2)$	&	AAbAbbBBaBaa	\\
$F_{14}(1,0.8,0.2)$	&	BabbbbBBBBAb	\\
$F_{15}(1,0.8,0.2)$	&	AbAbbbbBBBBaBa	\\
$F_{16}(1,0.8,0.2)$	&	AAbAbAAaaBaBaa	\\
$F_{17}(1,0.8,0.2)$	&	AAAbAbAaBaBaaa	\\
$F_{18}(1,0.8,0.2)$	&	BabbbbbBBBBBAb	\\
$F_{19}(1,0.8,0.2)$	&	AbAbbaBbABBaBa	\\
$F_{20}(1,0.8,0.2)$	&	AAAbAbbBBaBaaa	\\
$F_{21}(1,0.8,0.2)$	&	AbAbAbAbBaBaBaBa	\\
$F_{22}(1,0.8,0.2)$	&	AbAbAAAAaaaaBaBa	\\
$F_{23}(1,0.8,0.2)$	&	AAAbAbAAaaBaBaaa	\\
$F_{24}(1,0.8,0.2)$	&	AAAbAbbbBBBaBaaa	\\
$F_{25}(1,0.8,0.2)$	&	AAbAbAAAAaaaaBaBaa	\\
$F_{26}(1,0.8,0.2)$	&	AbAbAAAAAaaaaaBaBa	\\
$F_{27}(1,0.8,0.2)$	&	AAAAbAbbbbBBBBaBaaaa	\\
$F_{28}(1,0.8,0.2)$	&	AbAbBaBaabAaBAAbAbBaBa	\\
$F_{29}(1,0.8,0.2)$	&	AAAAAbAbbbbBBBBaBaaaaa	\\
$F_{30}(1,0.8,0.2)$	&	AbAbBaBAbABaAbaBabAbBaBa	\\
$F_{31}(1,0.8,0.2)$	&	AbAbBaBaBaBaAbAbAbAbBaBa	\\
$F_{32}(1,0.8,0.2)$	&	AAAAAbAbbbbbBBBBBaBaaaaa	\\
		\hline
	\end{tabular}
\end{table}

\begin{table}
	\centering
	\caption{The free group elements for the periodic three-body orbits}
	\label{table-A10}
	\begin{tabular}{ll} 
		\hline
$F_{i}(m_1, m_2, m_3)$ & free group element\\
		\hline
$F_{1}(0.6,0.8,1)$	&	BaAb	\\
$F_{2}(0.6,0.8,1)$	&	AbAaBa	\\
$F_{3}(0.6,0.8,1)$	&	BabBAb	\\
$F_{4}(0.6,0.8,1)$	&	AbAbBaBa	\\
$F_{5}(0.6,0.8,1)$	&	BabbBBAb	\\
$F_{6}(0.6,0.8,1)$	&	AbAbAaBaBa	\\
$F_{7}(0.6,0.8,1)$	&	BaBaaAAbAb	\\
$F_{8}(0.6,0.8,1)$	&	AbAbbBBaBa	\\
$F_{9}(0.6,0.8,1)$	&	AbABaAbaBa	\\
$F_{10}(0.6,0.8,1)$	&	BaBaaaAAAbAb	\\
$F_{11}(0.6,0.8,1)$	&	AbAbbbBBBaBa	\\
$F_{12}(0.6,0.8,1)$	&	AAbAbAaBaBaa	\\
$F_{13}(0.6,0.8,1)$	&	AAbAbbBBaBaa	\\
$F_{14}(0.6,0.8,1)$	&	BabbbbBBBBAb	\\
$F_{15}(0.6,0.8,1)$	&	AbAbbbbBBBBaBa	\\
$F_{16}(0.6,0.8,1)$	&	AAbAbAAaaBaBaa	\\
$F_{17}(0.6,0.8,1)$	&	AAAbAbAaBaBaaa	\\
$F_{18}(0.6,0.8,1)$	&	BabbbbbBBBBBAb	\\
$F_{19}(0.6,0.8,1)$	&	AbAbbaBbABBaBa	\\
$F_{20}(0.6,0.8,1)$	&	AAAbAbbBBaBaaa	\\
$F_{21}(0.6,0.8,1)$	&	AbAbAbAbBaBaBaBa	\\
$F_{22}(0.6,0.8,1)$	&	AbAbAAAAaaaaBaBa	\\
$F_{23}(0.6,0.8,1)$	&	AAAbAbAAaaBaBaaa	\\
$F_{24}(0.6,0.8,1)$	&	AAAbAbbbBBBaBaaa	\\
$F_{25}(0.6,0.8,1)$	&	AAbAbAAAAaaaaBaBaa	\\
$F_{26}(0.6,0.8,1)$	&	AbAbAAAAAaaaaaBaBa	\\
$F_{27}(0.6,0.8,1)$	&	AAAAbAbbbbBBBBaBaaaa	\\
$F_{28}(0.6,0.8,1)$	&	AbAbBaBaabAaBAAbAbBaBa	\\
$F_{29}(0.6,0.8,1)$	&	AAAAAbAbbbbBBBBaBaaaaa	\\
$F_{30}(0.6,0.8,1)$	&	AbAbBaBAbABaAbaBabAbBaBa	\\
		\hline
	\end{tabular}
\end{table}

\end{document}